\documentclass{article}

\usepackage{graphicx}
\usepackage{amssymb}
\usepackage{amsmath}
\usepackage{amscd}
\usepackage{pslatex}
\usepackage{apalike}

\usepackage{comment}

\title{A Link Diagram Visualizing Relations between Two Ordered Sets}
\author{T. Mizuno \thanks{Meijo University, eMail: tmizuno@meijo-u.ac.jp}}

\begin{document}

\maketitle

\abstract{%
This article provides a link diagram to visualize relations between two ordered sets representing precedences on decision-making options or solutions to strategic form games.
The diagram consists of floating loops whose any two loops cross just twice each other.
As problems formulated by relations between two ordered sets, I give two examples: visualizing rankings from pairwise comparisons on the diagram and finding Pareto optimal solutions to a game of prisoners' dilemma.
At visualizing rankings, we can see whether a ranking satisfies Condorcet's principle or not by checking whether the top loop is splittable or not.
And at finding solutions to the game, when a solution of the game of prisoners' dilemma is Pareto optimal, the loop corresponding to the solution has no splittable loop above it.
Throughout the article, I point out that checking the splittability of loops is an essence.
I also mention that the diagram can visualize natural transformations between two functors on free construction categories.
}


\section{Introduction}

Order is the essential structure of mathematical models.
We often draw directed graphs to visualize ordered sets.
Each node of the graphs represents an option of decision-making problems or a choice of strategic games, and each directed arc between a pair of nodes represents which node precedes another node.

However, drawing nodes and arcs is not suitable to represent local relations and global relations between two ordered sets.
Local relations are orders among elements in each set.
Global relations are correspondences between elements of a set and elements of another set.
When we use directed graphs, we will use arcs to represent the two relations; it makes us confused.
The two relations need two kinds of notations, respectively.

This article provides a link diagram that can represent local relations and global relations between two ordered sets.
By using the diagram, we can see that rankings from pairwise comparisons whether the rankings satisfy Condorcet's principle or not.
Then I describe how to represent games of the prisoners' dilemma on the link diagram.
We can find Pareto optimal solutions to the games with the diagram.
They are two types of abstract data visualization that represent relations between two ordered sets.
The former represents how an ordered set changes to another, and the latter represents how two people see a set.
The diagram also can represent a natural transformation among functors of categories.

\paragraph{Notice}
In this article, I write $x\succ y$ when an element $x$ precedes another element $y$, or $x$ is preferred to $y$, or $x$ beats $y$.

Generally, elements of ordered sets may have the following properties:
\begin{itemize}
\item (antisymmetry) for any elements $x$ and $y$, $x\succ y$ and $y\succ x$ iff $x=y$.
\item (comparability) for every pair of elements $x$ and $y$, $x\succ y$ or $y\succ x$.
\item (transitivity) for any elements $x$, $y$, and $z$, if $x\succ y$ and $y\succ z$ then $x\succ z$.
\end{itemize}
Ordered sets treated in this article have antisymmetry and comparability but may not have transitivity.

\section{A Link Diagram}
\label{sec:link-diagram}

\begin{figure}
\centering
\includegraphics[width=.5\linewidth]{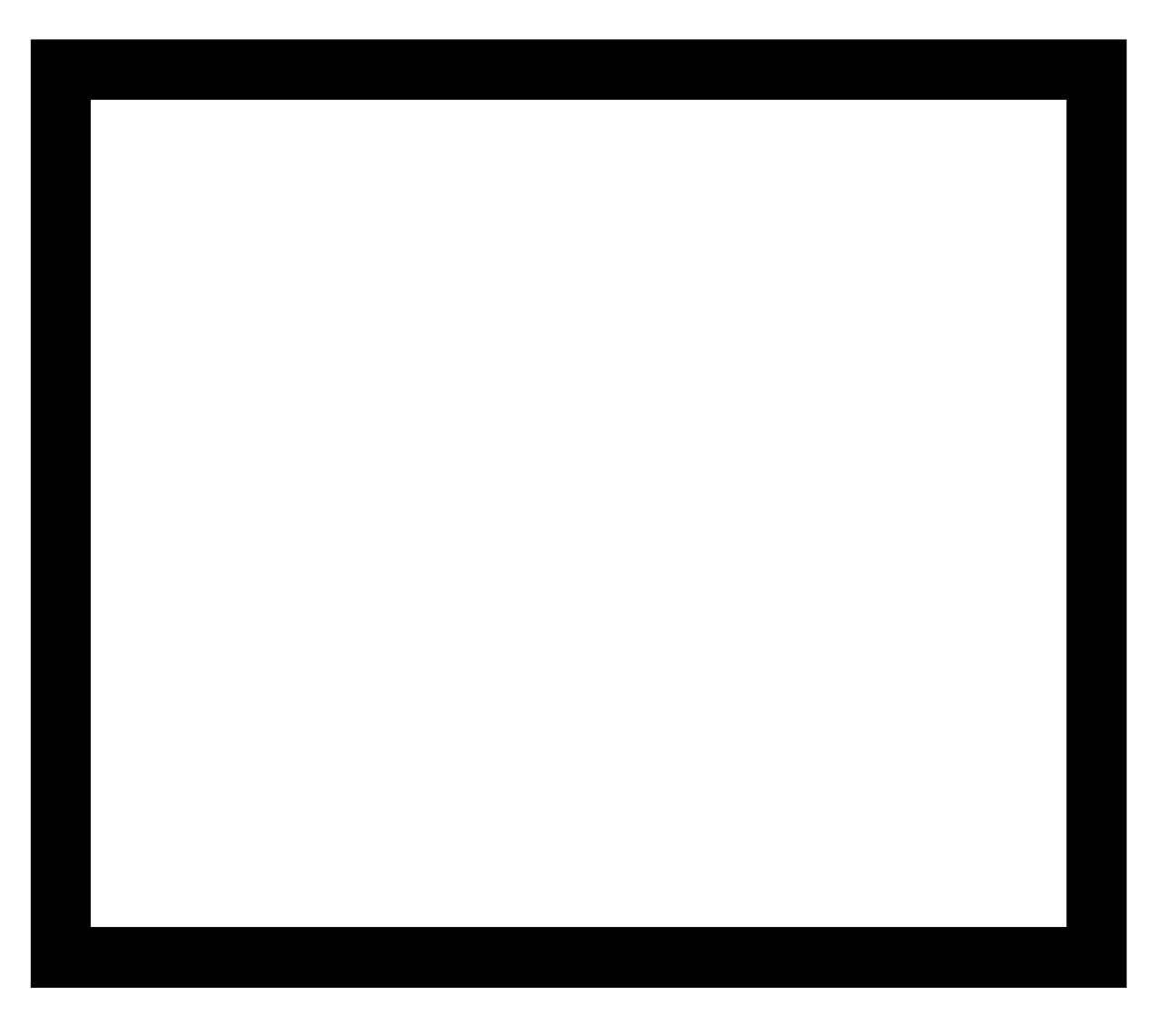}
\caption{%
A loop of the link diagram.
}
\label{fig:loop}
\end{figure}

A link diagram provided in this article consists of floating loops.
Each loop, such as Figure \ref{fig:loop}, represents an element of a set.
%
\begin{figure}
\centering
\includegraphics[width=.5\linewidth]{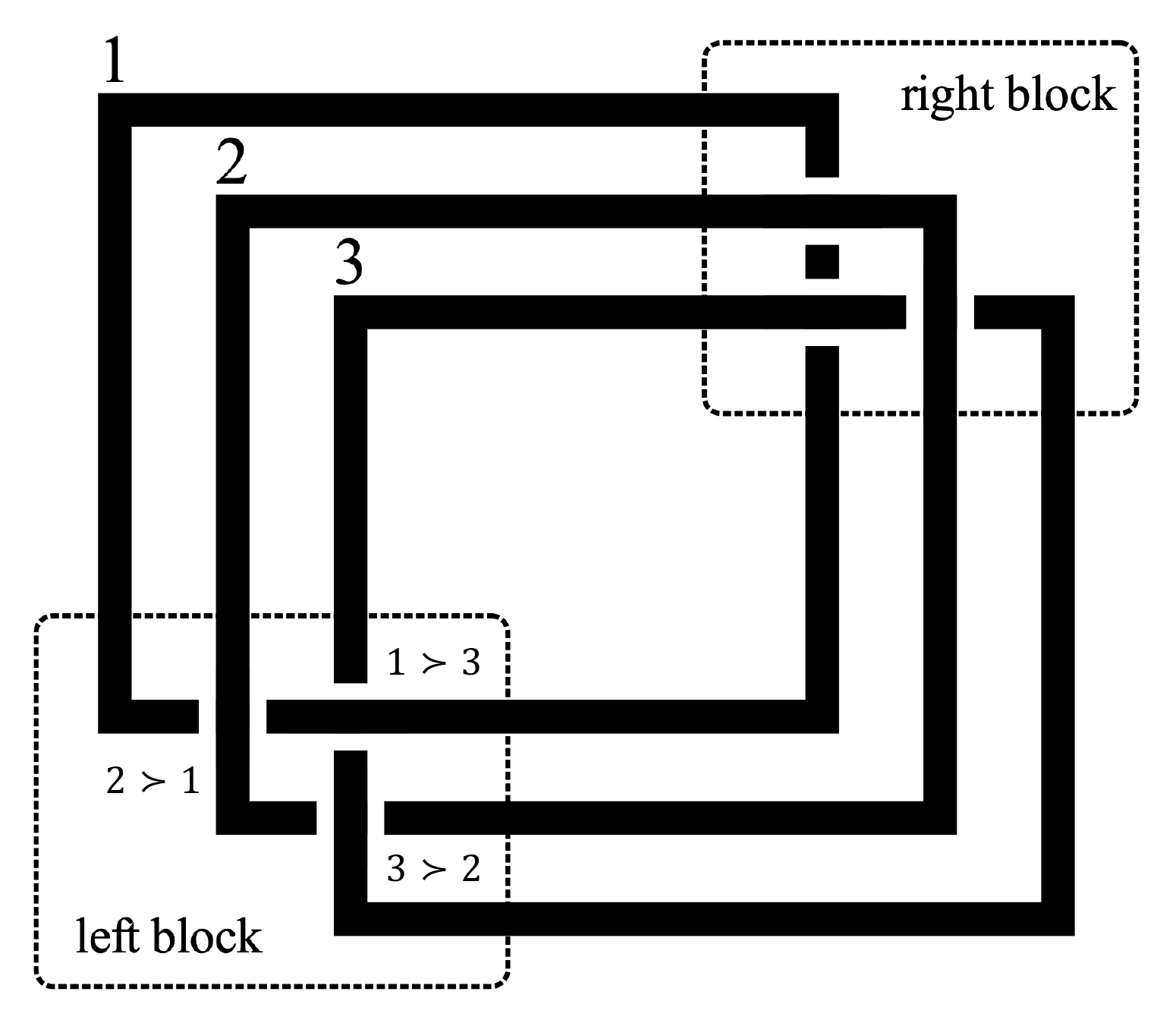}
\caption{%
A link diagram. The left block represents a cycle of preferences, and the right represents a total order.
}
\label{fig:ex}
\end{figure}

\begin{figure}
\centering\includegraphics[width=.5\linewidth]{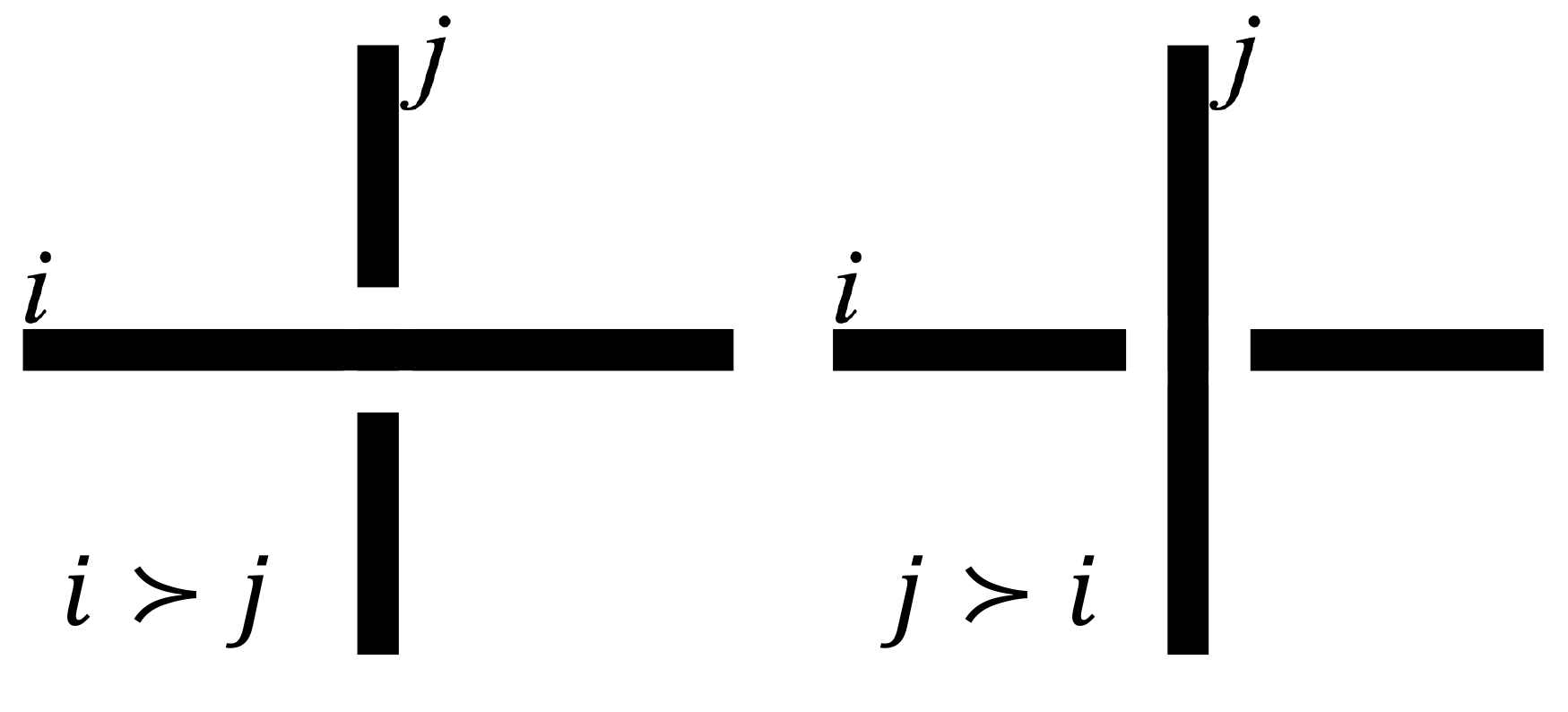}
\caption{%
Which loop is above means which element precedes.
}
\label{fig:cross-means}
\end{figure}

Figure \ref{fig:ex} is an example of link diagram.
Loops are arranged as every pair of loops intersect just two times on the left and the right sides of the diagram.
Each block on each side consists of crosses of loops represents an ordered set; the diagram represents two ordered sets.
How loops cross in each block represents local relations, which are precedence relations between elements in each order set.
When an element $i$ precedes $j$ in a set, the loop $i$ is above the loop $j$ at their intersection in the diagram (Figure \ref{fig:cross-means}).

In the diagram in Figure \ref{fig:ex}, there are three elements $1$, $2$, and $3$.
Crosses on the left block represent a cycles of preference; $2\succ 1$, $1\succ 3$, and $3\succ 2$.
While crosses on the right block represents a total order; $2\succ 3\succ 1$.

\section{Visualizing Ranking from Pairwise Comparisons}
\label{sec:ranking}
We often face to decide priorities among options in social choices, in private shoppings, and so on.
When we are hard to give ranks whole options at once, we will try pairwise comparisons.
For every pair of options, we compare them and declare which option is preferable to another, and then a ranking is aggregated from the comparisons.

There are many ways to aggregate the comparisons.
In this section, I deal with the geometric mean method that aggregate pairwise comparisons.
The method is used in AHP (the Analytic hierarchy process)  \cite{Saaty1980} which is a decision-making process.
AHP consists of three phases: 
construction of hierarchy structure, 
weighting options in each criterion,
and synthesis by the sum of products multiplied the weights by criteria's importance.
In the second phase, weighting options, pairwise comparisons are used in AHP.
For each criterion, decision-makers regard the criterion and show how an option is preferable to another one for every pair of options.
The preference is represented in a ratio greater than zero.

For example, in the case of three options $\{1,2,3\}$, in a criterion, ratios are shown by decision-makers and arranged into the following pairwise comparison matrix.
\begin{align}
    A = (a_{ij})=
    \begin{bmatrix}
        1 & 2 & 2\\
        \frac{1}{2} & 1 & 9\\
        \frac{1}{2} & \frac{1}{9}& 1
    \end{bmatrix}
    .
    \label{eq:ex-pcm3}
\end{align}
The element $a_{23}=9$ means that the option $2$ is $9$ times preferred to the option $3$.
If the option $i$ is preferred to $j$, $i\succ j$, then $a_{ij}>1$, otherwise $0< a_{ij}\leq 1$.

The geometric mean method calculates weight of option $i$ as the geometric mean of elements in $i$-th row of the matrix, and the weights in the example are $(w_{1}, w_{2}, w_{3})=(\sqrt[3]{4}, \sqrt[3]{9/2}, \sqrt[3]{1/18})$, where $w_{i}$ is the weight of $i$.
It means $2\succ 1 \succ 3$, and the winner is the option $2$.
\footnote{%
The principal eigenvector method is also widely used to calculate weights from pairwise comparisons in AHP.
In three options cases, its result is equivalent to the geometric mean method by ignoring positive constant multiple.
}

Now, let us look again at the pairwise comparisons.
The option $1$ precedes all of the others; $1\succ 2$ ($a_{12}=2>1)$ and $1\succ 3$ ($a_{13}=2>1$).
That is quite unnatural.
If the option $i$ wins $j$ in the result of aggregation, then we expect that $i$ precedes $j$ in their pairwise comparison.
The example shows that the geometric mean method may not satisfy the requirement.

In the 18th century, Condorcet mentioned the requirement as now known as Condorcet's principle \cite{Condorcet1785}.
He claimed that an option that beats all of the others in pairwise comparisons must be the winner in the result of aggregation when such option exists.

I introduce {\it naturality} that is a modified condition of Condorcet's principle.
Naturality is the condition that an option $i$ beating $j$ in aggregation result must precede $j$ directly or indirectly in their pairwise comparison.

An option beating all others has no option preceding it either directly or indirectly.
So, if naturality is satisfied, Condorcet's principle is satisfied.
And more, an option beaten by all of the others in pairwise comparisons is the lowest in ranking satisfying naturality when such option exists.

\begin{figure}
\centering
\includegraphics[width=.5\linewidth]{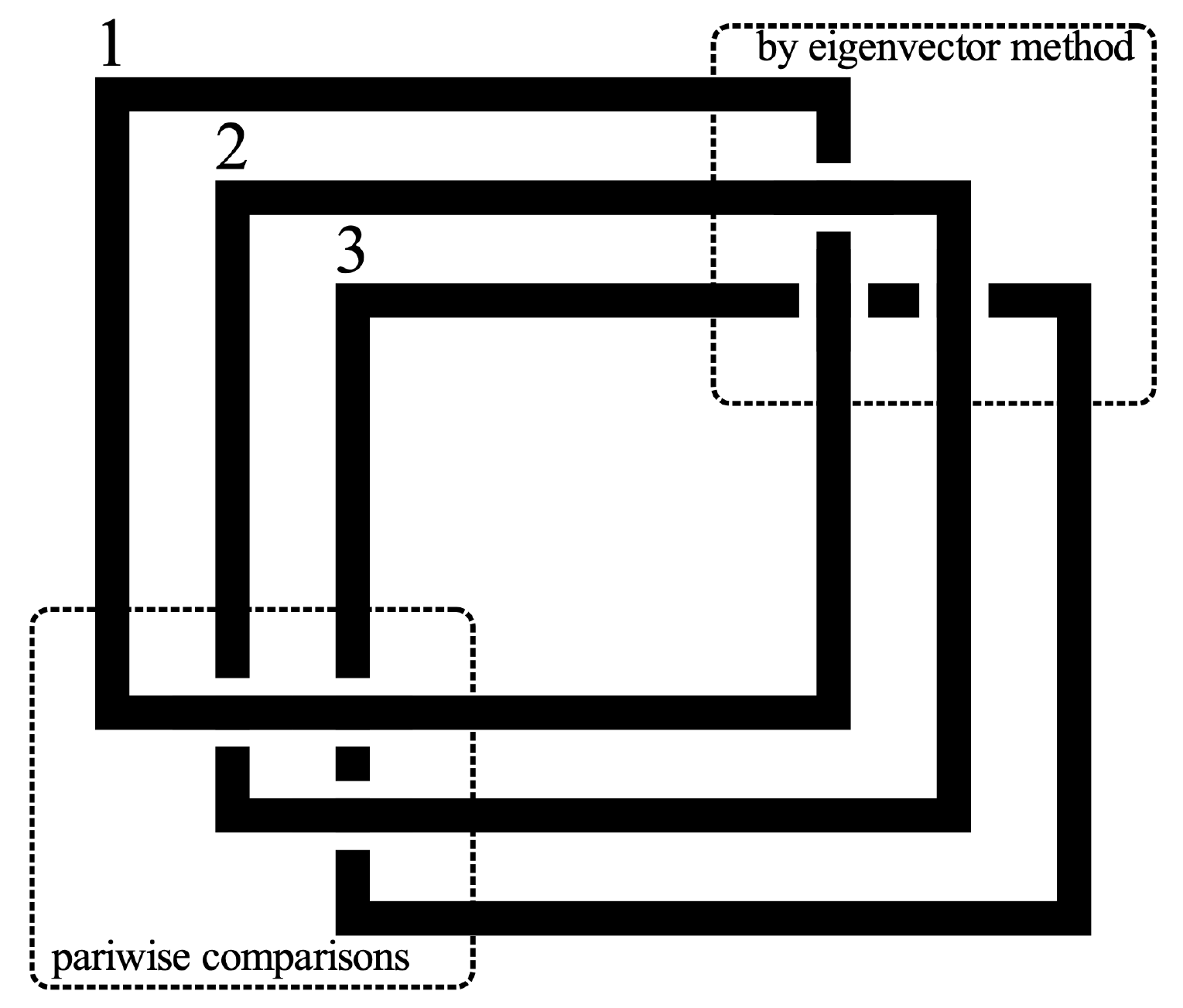}
\caption{%
A ranking whose pairwise comparison matrix is (\ref{eq:ex-pcm3}), and whose aggregation by the geometric mean method $2\succ 1\succ 3$.
}
\label{fig:ranking-by-gm}
\end{figure}

\begin{figure}
\centering
\includegraphics[width=.5\linewidth]{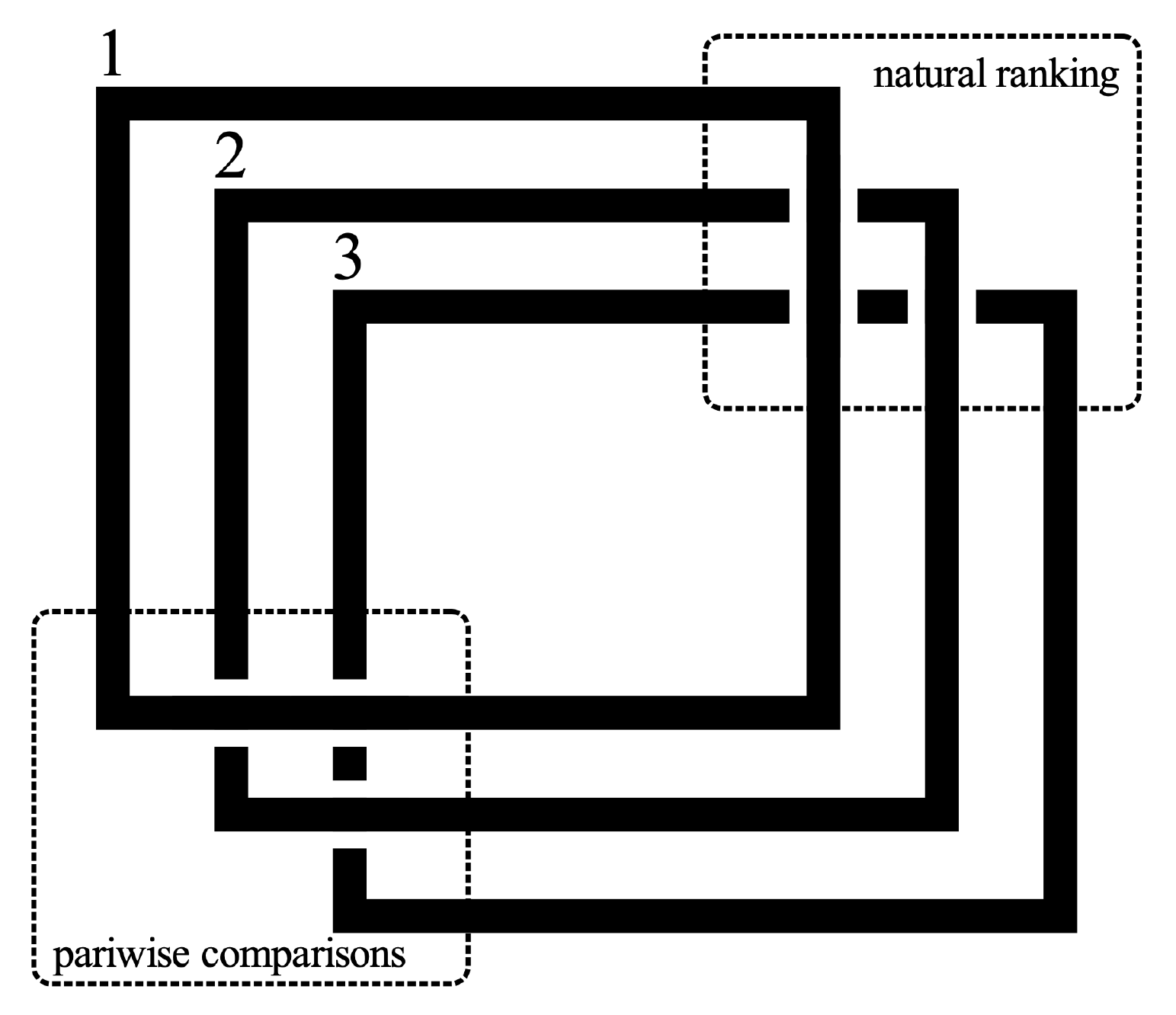}
\caption{%
A ranking that satisfies naturality. Its pairwise comparison matrix is (\ref{eq:ex-pcm3}), and its aggregation is $1\succ 2\succ 3$. The top loop and the bottom loop are splittable.
}
\label{fig:natural-ranking}
\end{figure}

We can easily see whether a ranking satisfies the naturality by visualizing the ranking on the link diagram.
Let us put the ordered set representing pairwise comparisons on the left block and put the ordered set of the aggregation result on the right.
The ranking of the above example aggregated by the geometric mean method is in Figure \ref{fig:ranking-by-gm}.
While a ranking satisfying naturality for the example is in Figure \ref{fig:natural-ranking}.

If an option beats all other options in pairwise comparisons, its loop is the top on the link diagram's left block.
It is clear that if such an option exists and the ranking satisfying naturality, its loop is splittable from all other loops.
Similarly, if an option beaten from all other options exists and the ranking satisfying naturally, its loop is splittable from all other loops.
So, we can check whether a ranking satisfies naturality or not is whether loops are splittable or not in the link diagram.

Then, you may have a question: Are there such a ranking way always?
The answer is yes.
An ordered set treated in this article is equivalent to a tournament graph.
A tournament graph is a kind of directed graph that must have at least one Hamilton path \cite{Wilson2012}.
The path is a directed path that visits all nodes exactly once.
That is, extracting the path is the way of ranking satisfying naturality.

\section{Visualizing The Game of Prisoners' Dilemma}
\label{sec:game}
The game of prisoners' dilemma is a model of interaction of multiple players whose choices affect each other's outcomes.

Let us consider a criminal trial in that two prisoners face a choice to confess their crime or keep silent.
If they both keep silent, their both imprisonments will be short term, two years, because of insufficient evidence of the crime.
If only one of them confesses, his/her imprisonment will be reduced to one year, and the confession will be a witness against the other; the other's imprisonment will be five years.
If they both confess, their imprisonment will be four years less than five years because of cooperating with the authorities.

We can model the criminal trial as a strategic game.
Two prisoners are player A and player B, and they choose simultaneously one from their strategies: to confess or keep silent.
Outcomes of players A and B by their choices are in Table \ref{table:outcomes-of-A} and Table \ref{table:outcomes-of-B}, respectively.
\begin{table}
  \centering
    \caption{Outcoms of palyer A}
    \label{table:outcomes-of-A}
  \begin{tabular}{|c||c|c|}
    \hline
    A's imprisonment&B confesses & B keeps silent\\ \hline\hline
    A confesses & 4 years & 1 year\\ \hline
    A keeps silent & 5 years & 2 years\\ \hline
  \end{tabular}
  \label{tb:mulrow}
\end{table}
\begin{table}
  \centering
    \caption{Outcoms of player B}
    \label{table:outcomes-of-B}
  \begin{tabular}{|c||c|c|}
    \hline
    B's imprisonment &B confesses & B keeps silent\\ \hline\hline
    A confesses & 4 years & 5 years\\ \hline
    A keeps silent & 1 year & 2 years\\ \hline
  \end{tabular}
  \label{tb:mulrow}
\end{table}

A pair of their choices are referred to as a solution to the game.
By using notations C and S for confession and for keeping silent, respectively, there are four solutions CC, CS, SC, and SS.
For example, CS means player A confesses, and player B keeps silent.
Players, of course, prefer shorter imprisonment to longer, so player A's precedence of the solutions is CS$\succ$SS$\succ$CC$\succ$SC, while player B's is SC$\succ$SS$\succ$CC$\succ$CS.

\begin{figure}
\centering
\includegraphics[width=.5\linewidth]{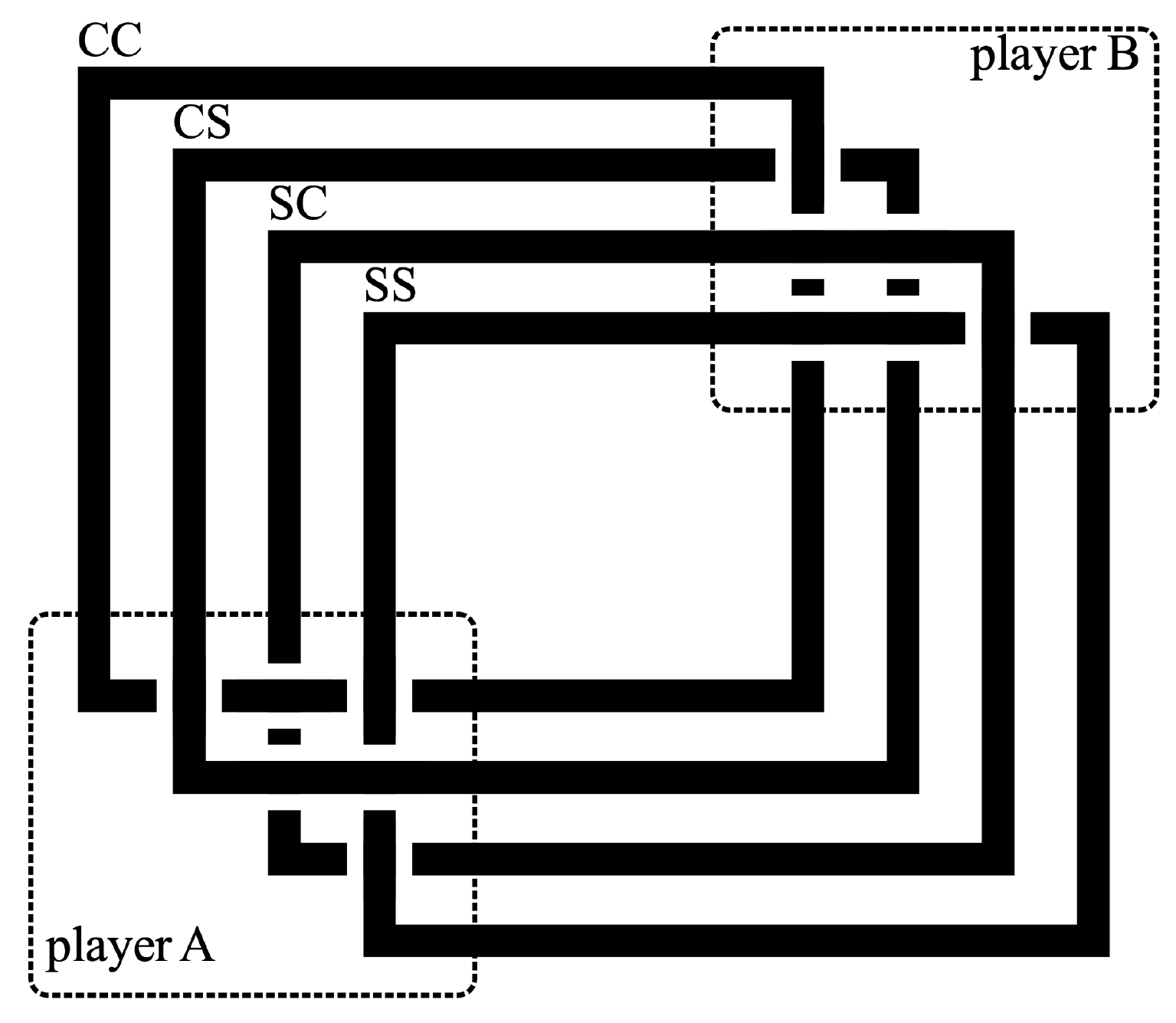}
\caption{%
A link diagram that represents the game of prisoners' dilemma of Table \ref{table:outcomes-of-A} and Table \ref{table:outcomes-of-B}.
}
\label{fig:ex-prisoners-dilemma}
\end{figure}

We can represent the game on a link diagram by arranging player A's precedence in the diagram's left and player B's in its right (Figure \ref{fig:ex-prisoners-dilemma}).

In such strategic games, which solution must players choose?
There are some concepts to decide on the choices.
In this article, I focus on Pareto optimality.

If one player takes a strategy, and no strategy shrinks his/her imprisonment without extending other's imprisonment, and another player also takes such strategy, then the solution that consists of the strategies is Pareto optimal.
In other words, if they can shrink both imprisonments by changing both strategies from a solution, the solution is not Pareto optimal.

\begin{figure}
\centering
\includegraphics[width=.5\linewidth]{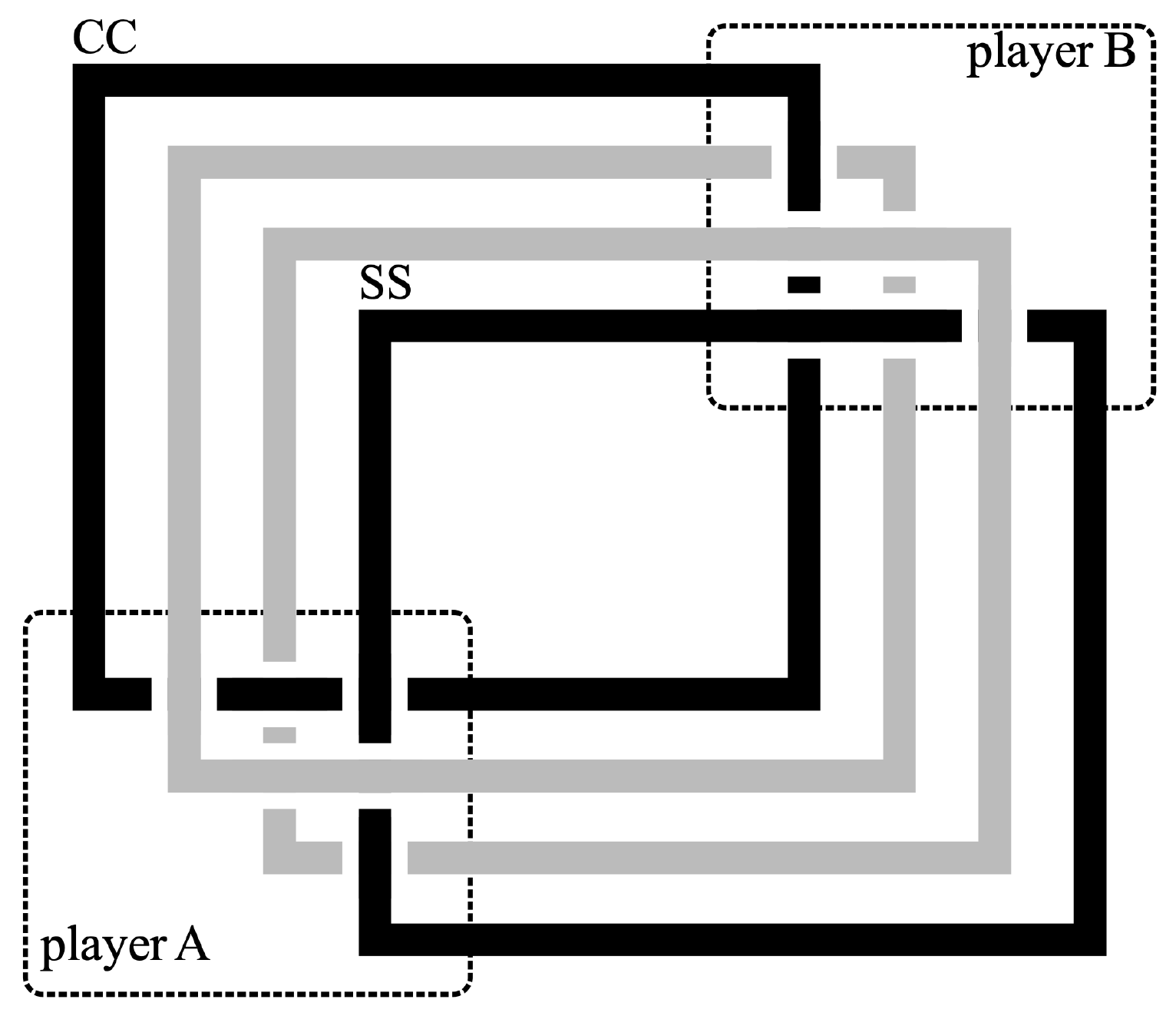}
\caption{%
The loop CC and SS are splittable from each other.
}
\label{fig:ex-CC-and-SS}
\end{figure}

In the example, the solution CC is not Pareto optimal.
Both players can shrink their imprisonments by changing their strategies from C to S.
Existing the loop SS above CC represents it, and both loops are splittable from each other (Figure \ref{fig:ex-CC-and-SS}).

In the link diagram, a Pareto optimal solution has no splittable loop above it; CS, SC, and SS are Pareto optimal.

\section{Discussion}
\label{sec:discussion}

We can build a free construction category from an ordered set \cite{Leinster2014}.
Category is an abstract mathematical structure that consists of objects and arrows.
A category is built from an ordered set as follows.
Each object of the category corresponds to each element of the ordered set.
If an element $i$ precedes $j$, $i\succ j$, in the set, then an arrow $i \rightarrow j$ exists in the category.
Composited arrows are also in the category.
An arrow $i \rightarrow j$ and an arrow $k \rightarrow l$ can be composited as $i \rightarrow j \rightarrow l$ when $j=k$.
Category made as above is often referred to as the free construction category.

Under the construction, the existence of an arrow from $i$ to $j$ means that the element $i$ precedes $j$ directly or indirectly.
If the element $i$ precedes $j$ directly, then $i\rightarrow j$ exists.
If $i$ precedes $j$ indirectly, then there is a composited arrow from $i$ to $j$.
If there is the composited arrow is $i\rightarrow p\rightarrow q\rightarrow j$, then $i$ may be defeated by $j$ directly, but $i$ beats $j$ indirectly; $i\succ p$ and $p\succ q$ and $q\succ j$.

\begin{figure}
\centering
\includegraphics[width=.5\linewidth]{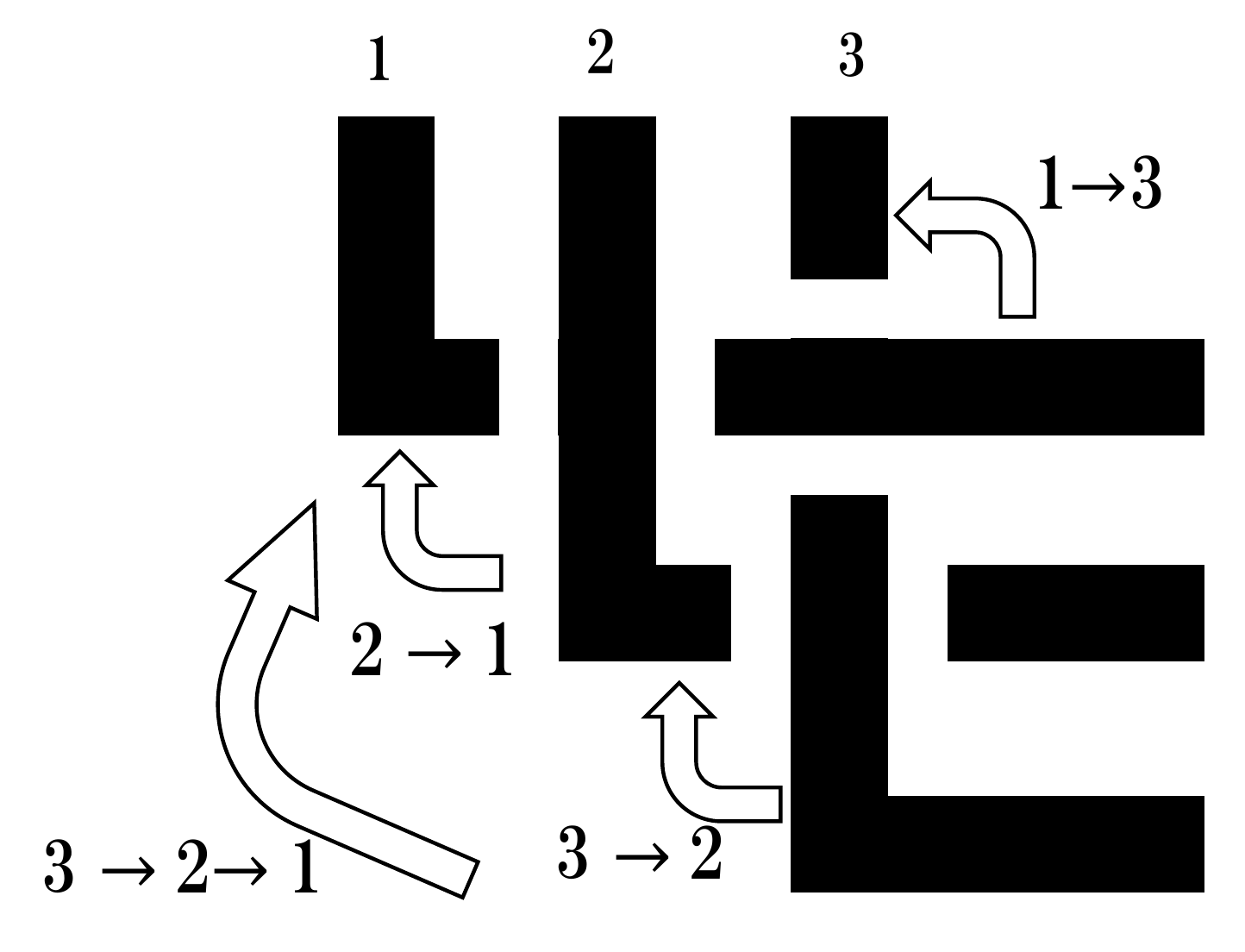}
\caption{%
Examples of arrows and a composited arrow in the link diagram.
}
\label{fig:jump}
\end{figure}

In the link diagram, an arrow $i\rightarrow j$ of the category is represented as jump from the upper loop $i$ to the lower loop $j$ at crossing of the two loops.
In Figure \ref{fig:jump}, the arrow $1\rightarrow 3$ represents jump from the loop of $1$ to $3$ at the crossing the loop of $1$ and $3$.
It can jump from $3$ to $2$ and $2$ to $1$ in the diagram.
It means the category has the composited arrow $3\rightarrow 2\rightarrow 1$.

In the example of ranking from pairwise comparisons in section \ref{sec:ranking}, categories represent a set of pairwise comparisons and a total order sets.
Let us denote the former category as $\mathcal{L}$ (local), the latter category as $\mathcal{G}$ (global), and a correspondence, referred to as functor, from $\mathcal{G}$ to $\mathcal{L}$ as $F$.
Categories in this article, objects are same in both category; $F(i)=i$, where $i$ is an object of $\mathcal{G}$.
In particularly, the functor is called covariant functor when an arrow $i\rightarrow j$ is corresponded to a arrow from $i$ to $j$; $F(i\rightarrow j)=i\rightarrow \cdots \rightarrow j$.
So, ranking satisfies naturality is building a category $\mathcal{G}$ and a covariant functor $F$ from $\mathcal{G}$ to $\mathcal{L}$.

And also, naturality is represented as a commutative diagram of a natural transformation $\eta$ from the identity functor $id$, which changes nothing, to the covariant functor $F$.
\begin{align}
    \begin{CD}
   F(i)=i @<{\eta}<< i=id(i)\\
@V{F(i\rightarrow j)=i\rightarrow\cdots\rightarrow j}VV    @VV{i\rightarrow j=id(i\rightarrow j)}V \\
   F(j)=j   @<{\eta}<<  j=id(j)
\end{CD}
\end{align}
It means that the link diagram represents a natural transformation.


\section{Conclusions}
\label{sec:conclusions}
In this article, I provided a link diagram to visualize a relation between two ordered sets.

I demonstrated that the link diagram could visualize rankings from pairwise comparisons.
We can grasp whether a ranking satisfies naturality, a generalization of Condorcet's principle, or not by checking whether the top loop of the diagram is splittable or not.

I also visualized a game of prisoners' dilemma with the diagram.
I pointed out that a solution to the game is Pareto optimal when the solution's loop does not have splittable loops above it.
Of course, the link diagram can represent other strategic form games such as the battle of sexes and matching pennies.

After them, I mentioned that the ordered sets are considered categories, and the link diagram also represents categories and, in particular cases, their natural transformations.

In the field of knot theory, the diagram in this article is a link whose arbitrary two loops have just two intersections.
Splitability of the link can represents two concepts: naturality and Pareto optimality.
I am searching other usefull concepts visualized on the link.

\bibliographystyle{apalike}

\end{document}